\documentstyle[preprint,eqsecnum,aps]{revtex}

\begin{document}
\draft
\tightenlines
\preprint{
\rightline{\vbox{\hbox{\rightline{MSUCL-934}}
\hbox{\rightline{McGill/94-44}}}}
         }
\title{Higher excitations of $\bbox{\omega}$ and
$\bbox{\phi}$ in dilepton spectra}
\author{Kevin Haglin\cite{myemail}}
\address{
National Superconducting Cyclotron Laboratory, Michigan
State University\\
East Lansing, Michigan 48824--1321
}
\author{Charles Gale\cite{charlesemail}}
\address{
Physics Department, McGill University, 3600 University St., Montr\'eal,
Qu\'ebec, H3A 2T8, Canada
}
\date{\today}
\maketitle
\begin{abstract}
We consider lepton pair production via two-hadron annihilation
through various isoscalar vector mesons within
hot, baryon-free matter.  This is tantamount to constructing
effective form factors which we model using a vector-meson-dominance
approach and compare with experiment.
In particular, we consider the reactions
$\pi\rho\to e^+e^-$ and $\bar K K^{*}(892)$ + c.c.$\to e^+e^-$.
We find that $\omega(1390)$ and $\phi(1680)$ are
visible in the mass spectrum for the thermal production rate
above the $\pi^{+}\pi^{-}\to e^+e^-$ tail and even above
the $\pi a_{1}\to e^+e^-$ results---both of which were considered
important in their respective mass regions.
\end{abstract}

\narrowtext

\section{Introduction}
\label{sec:intro}

Production of dileptons in high-energy heavy-ion collisions continues
to be studied actively since it is hoped that such signals
might be useful in discriminating prehadronic or quantum chromodynamic
(QCD) plasma formation from ordinary hadronic matter.  The possible
success depends on having a reasonably complete picture
of both plasma and hadronic phases.  Crossover from one to the
other is expected for a critical temperature in the neighborhood of 200 MeV
and near this boundary the hadronic matter will be populated by many
species of strange and nonstrange mesons.  However, to get a rough idea about
this phase of matter one typically used a simple pion-gas approximation.
In terms of dilepton spectra, it has recently been shown that
thermal dielectron production rates from the pion gas could be an
order of magnitude lower than those from a hadron gas consisting of pions and
various resonances\cite{cgpl94}.  Going beyond simple pion gas considerations
is important.  Allowing heavier species to partially
comprise the system complicates matters since each species might
interact among themselves as well as with others.   The list of possible
processes then becomes quite lengthy.  Further complications or uncertainties
arise since, with the exception of the pion\cite{dm2coll},
hadronic form factors are not particularly well known.
Reactions which involve an intermediate isoscalar vector meson could
contribute to dilepton production and for this reason the mass distribution
of produced dileptons is expected to exhibit structure owing to
these hadronic resonances.

Observed spectra from heavy-ion collisions will of course be restricted to
a time-integrated yield.  Temperature (or time) dependent rates are the
essential quantity that might be input for models of
the time evolution in order to make predictions for overall yields.  Rate
calculations have been done with various levels of sophistication.
In a quest for more accurate knowledge of dielectron production
rates we consider some hadronic processes which produce
$e^+e^-$ pairs and then compare with spectra calculated
previously\cite{cgpl94,cscmkcg}.  The particular processes we
consider are pseudoscalar and vector reactions of the type
$P+V \rightarrow V^{I=0} \rightarrow e^+e^-$ where $P$ and $V$
are either pions and $\rho$-mesons or kaons and $K^*(892)$s.  The
intermediate isoscalar vector hadronic resonances are $\phi(1020),
\omega(1390)$ and $\omega(1600)$ for the $\pi\rho$ processes and
$\phi(1680)$ for the strange-particle process.  Special attention
is payed to experimental data on the time-reversed processes in order
to constrain or even fix the coupling constants.

Our paper is organized in the following way.  In Sect.~\ref{sec:theory}
we present the basic formalism for the interactions among the hadronic
degrees of freedom and the elementary scattering (annihilation) processes.
A vector-meson-dominance (VDM) approach will be presented which allows
calculation of the basic Feynman diagram for $a+b\rightarrow V^{I=0}\rightarrow
\gamma^{*} \rightarrow e^+e^-$.  Within this approach the amplitude
for each diagram contains a ratio of the strong-to-electromagnetic
coupling constants.  Experimental data on cross sections for the
time reversed processes $e^+e^-\rightarrow a+b$ are shown as compared
with model calculations.  These are used to fix coupling constants
and phases between diagrams.  One is really constructing an effective
form factor by this prescription.  Section ~\ref{sec:thermalrates}
contains the essential details and results for thermal production
rates which include effects from these hadronic resonances.  The
interesting result is that the $\omega(1390)$ and $\phi(1680)$
peaks are visible above the rate from $\pi^{+}\pi^{-}$ annihilation
and even $\pi a_{1}$ reactions which have recently been shown to be
dominant for restricted masses\cite{cscmkcg}.  Finally, in
Sect.~\ref{sec:conclusion}
we conclude with some final remarks which essentially stress
the importance of appealing to experimental data
at some level for the elementary reactions in future studies.  Putting them
into calculations on systems of hadronic matter only after
such comparison minimizes uncertainty.

\section{Formalism}
\label{sec:theory}

After the initial stage of a collision of the type we consider,
which may well include a prehadronic or quark-gluon-plasma (QGP)
phase, hadronization gives rise to a hot ensemble of the lighter
mesons.  Within the context of a thermodynamic model this system's
composition has recently been studied\cite{khsp94}.
The most abundant hadronic species in this scenario
are pions, kaons, $\rho$-mesons and $K^*(892)$ strange vector
mesons owing to the large spin-isospin degeneracy of the last two.  With
temperatures ranging from phase boundary or crossover temperature
downward toward the pion mass or lower, where the system ceases
interacting, the number densities of these species range from
several hundredths to a few tenths per fm$^{3}$\cite{khsp94}.  Until
freezeout, they interact very strongly.  Though binary reactions
dominate at these densities, studies of the effect of ternary
reactions have been pursued\cite{pl94}.  In this work
our aim was to study the relative strength of different lepton pair
producing sources and the possible manifestation of some higher-lying
resonances. For this, we have chosen a simple thermal
environment on purpose. As is well known from the analysis of heavy ion
collisions in the intermediate energy regime, nonequilibrium effects might
manifest themselves in the final state. A related issue here is the
importance of Drell-Yan and $D \bar{D}$ lepton pair signals around the
invariant mass region adjacent to the $J/\psi$\cite{NA38}.  Those are
obviously important and should be adressed in a realistic comparison with
heavy ion experimental data.

Species can be classified according to their quantum numbers in order
to identify the manner with which we model their interactions.  Pseudoscalar
particles (P) interact with a vector particle (V) in a manner described
by the Lagrangian\cite{um88}
\begin{equation}
L_{VPP} = g_{VPP}\, V^{\mu} P \stackrel{\leftrightarrow}
{\partial_{\mu}} P
\label{eq:lvpp}
\end{equation}
and vectors interact with a pseudoscalar through
\begin{equation}
L_{VVP} = g_{VVP}\,
\epsilon_{\mu\nu\alpha\beta}\, \partial^{\mu}V^{\nu}
\partial^{\alpha}V^{\beta}\,P.
\label{eq:lvvp}
\end{equation}

Let two hadrons, call them $a$ and $b$, scatter in
a timelike fashion, i.e. let them annihilate to form a resonant
state $V$---restricted here to neutral vector mesons.
It is not entirely clear how it converts itself from a hadronic
to an electromagnetic field, but we model the phenomenon using the principle
of vector-meson dominance.  In practice, it means that the neutral component
of the hadronic vector field couples directly to the photon via\cite{js60}
\begin{equation}
L_{VA} = \left({em_{V}^{2}\over g_{V}}\right) \, V^{(3)}_{\mu} A^{\mu}
\label{eq:lem}
\end{equation}
where $V^{(3)}$ is the neutral component of the resonant isovector field
and $A$ is the electromagnetic field.
Using Eqs.~(\ref{eq:lvpp}) or (\ref{eq:lvvp}) and (\ref{eq:lem}) we
construct the general annihilation diagram through a vector resonance
$V$ to a virtual or massive photon, which subsequently decays into a
lepton pair as shown in Fig~\ref{fig:abtoe+e-}.
This differs from an analogous process of purely electromagnetic
exchange because of the presence of the hadronic resonance.  Since the
effect of this resonance appears as a function of invariant mass alone,
the hadronic dependence of $V$ at the vertex factorizes from the rest of
the diagram in Fig.~\ref{fig:abtoe+e-} as a form factor times a pointlike
electromagnetic process.  Vector-meson dominance has been very successful
in the areas of hadronic form factors, photoproduction and absorption
cross sections, and in the vector meson exchange contributions to $\pi N$
and $NN$ scattering.

An invariant amplitude for scattering or decay in which the photon couples
to the vector meson contains the ratio $g_{Vab}/g_{V}$ which
forces the appearance of the same ratio (squared) in the cross section or
decay rate. We will discuss
two methods for fixing these quantities.  For the case
of the $\phi(1020)$ we can fix these coupling constants independently via
strong and electromagnetic decays; whereas for the
higher excitations presently considered we are forced to use another procedure.
First consider the decay $\phi \to \rho\pi$.  We calculate its
rate to be
\begin{equation}
\Gamma_{\phi\to \rho\pi} = {g_{\phi\rho\pi}^{2} |\bbox{p}\,|^{3}
\over 12\pi}
\label{eq:phirhopi}
\end{equation}
where $\bbox{p}$ is the center-of-mass momentum of the decay products.
This determines $m_{\phi}^{2}g_{\phi\rho\pi}^{2}/4\pi$ = 0.29 when
the partial decay rate is taken to be 12.9\% of 4.43 MeV.  The same
expression and numerical conclusions were reached in a recent study
of $\phi$-meson properties at zero and finite temperature\cite{khcg94}.
Next, the electromagnetic decay $\phi\to e^+e^-$ is modeled and computed
to be
\begin{equation}
\Gamma_{\phi\to e^+e^-} = {4\pi\alpha^{2}m_{\phi}\over 3 g_{\phi}^{2}}.
\label{eq:phie+e-}
\end{equation}
Using the branching fraction $\Gamma_{\phi\to e^+e^-}/
\Gamma_{\phi}^{\rm full} = 3.09\times 10^{-4}$, we arrive at the dimensionless
vector-dominance coupling constant for the $\phi$ of $g_{\phi}$ = 12.9.

Since data are not sufficient to allow us to follow this
prescription for the other resonances, we instead
appeal to the time-reversed hadron production processes
\begin{equation}
e^{+}e^{-} \to a+b+\ldots
\end{equation}
We compute the cross sections for $e^+e^-\to \rho\pi$ and for
$e^+e^-\to \bar KK^{*}(892)$ which is approximated by
$e^+e^-\to K_{S}^{0}K^{\pm}\pi^{\mp}$.  Henceforth $K^{*}(892)$
will be written just as $K^{*}$.
The diagrams are shown in Figs.~\ref{fig:timereversed}a and
\ref{fig:timereversed}b.
The cross section for the process in Fig.~\ref{fig:timereversed}a
is found to be
\begin{equation}
\sigma_{e^+e^-\to \pi\rho}(M) = {4\pi\alpha^{2}|\bbox{p}\,|^{3}\,
|F_{\pi\rho}(M)|^{2} \over 3 M^{3}}
\end{equation}
where $M=\sqrt{\bbox{p}^{2}+m_{\pi}^{2}}+\sqrt{\bbox{p}^{2}+m_{\rho}^{2}}$
and
\begin{equation}
F_{\pi\rho}(M) = \sum\limits_{V}\left({g_{V\pi\rho}\over g_{V}}\right)
{e^{i\varphi_{V}}m_{V}^{2}\over m_{V}^{2}-M^{2}-im_{V}\Gamma_{V}},
\label{eq:ffpirho}
\end{equation}
with the sum running over the three vectors $\phi(1020), \omega(1390)$
and $\omega(1600)$.  $\omega(783)$ could also be included in the sum
but since it is so far off shell for masses considered here
it can safely be neglected.
Comparison with experiment constrains the ratios of coupling constants
as well as phase angles in Eq.~(\ref{eq:ffpirho}).
The comparison to data is shown in
Fig.~\ref{fig:e+e-topirho}.  Details of the parameters in our model
go as follows.  For the phases we take $\varphi_{\phi}=0$ (without
loss of generality one of them can be set to zero),
$\varphi_{\omega}=5\pi/4$ and $\varphi_{\omega^{\prime}}=\pi$; while
the ratios are taken to be
$g_{\omega\pi\rho}/g_{\omega} = 0.29$ GeV$^{-1}$ and
$g_{\omega^{\prime}\pi\rho}/g_{\omega^{\prime}} = 0.05$ GeV$^{-1}$.
Note that $g_{\phi\pi\rho}/g_{\phi} = 0.15$ GeV$^{-1}$ was already
fixed by Eqs~(\ref{eq:phirhopi}--\ref{eq:phie+e-}).
Notation used here is that $\phi$ means $\phi(1020)$,
$\omega$ stands for $\omega(1390)$ and
$\omega^{\prime}$ is $\omega(1600)$.

The strange particle process $e^+e^-\rightarrow \bar{K}K^{*}$
approximated by the reaction with the eventual three-body final
state is shown in Fig.~\ref{fig:timereversed}b.  Since the decay of
$K^{*}$ is practically entirely into $\pi+K$, these are nearly equivalent.
We take a relativistic Breit-Wigner form for the cross section.
Computing $\Gamma_{e^+e^-\to \phi^{\prime}}$ and
$\Gamma_{\phi^{\prime}\to \bar{K}K^{*}}$, $\sigma$ can be written as
\begin{equation}
\sigma_{e^+e^-\to \bar{K}K^{*}}(M) = {\pi\alpha^{2}
\lambda^{3/2}(M^{2},m_{\bar{K}}^{2},m_{K^{*}}^{2})\over 8M^{4}
m_{\phi^{\prime}}^{2}} \left|F_{KK^{*}}(M)\right|^{2}
\end{equation}
where $\lambda(x,y,z) = x^{2}-2x(y+z)+(y-z)^{2}$ and
\begin{equation}
F_{KK^{*}}(M) = \left(g_{\phi^{\prime}KK^{*}}\over g_{\phi^{\prime}} \right)
{ m_{\phi^{\prime}}^{2}
\over m_{\phi^{\prime}}^{2}-M^{2}-im_{\phi^{\prime}}\Gamma_{\phi^{\prime}}}.
\label{eq:ffkkstar}
\end{equation}
Using value $g_{\phi^{\prime}KK^{*}}/g_{\phi^{\prime}}$ = 0.19 GeV$^{-1}$
we generate the curve shown in Fig.~\ref{fig:e+e-tokkpi}
compared with experiment.  Also shown is the result from a
fully microscopic calculation of the exact
diagram of Fig.~\ref{fig:timereversed}b whose cross section involves
integration over full three-body phase space.
Given the forms of the interactions written in
Eqs.~(\ref{eq:lvpp}) and (\ref{eq:lvvp}),
the model form of the cross section is well defined.
It is presented as the dashed curve.  The same ratio of coupling
constants is used in both approaches.

\section{Thermal Production Rates}
\label{sec:thermalrates}

Hadrons in this hot reaction zone are assumed to have their momenta
thermally distributed.  Then the differential rate at which they
scatter, i.e. the differential
rate at which hadrons annihilate to create an $e^+e^-$ pair of
invariant mass $M$ can be written as
\begin{eqnarray}
{dR\over dM^{2}} &=& {\cal N} \int
{d^{3}p_{a}\over 2E_{a}(2\pi)^{3}}
{d^{3}p_{b}\over 2E_{b}(2\pi)^{3}}
{d^{3}p_{+}\over 2E_{+}(2\pi)^{3}}
{d^{3}p_{-}\over 2E_{-}(2\pi)^{3}}f(E_{a})f(E_{b})\nonumber\\
& &\quad\times (2\pi)^{4}
|\,\overline{\cal M}\,|^{2} \delta^{4}(p_{a}+p_{b}-p_{+}-p_{-})
\delta\left(M^{2}
-(p_{+}+p_{-})^{2}\right)
\label{eq:diffrate}
\end{eqnarray}
where $f$ is the Bose-Einstein distribution and ${\cal N}$ is
an overall degeneracy factor.  With the aid of the $\delta$ functions
some of the phase space can be reduced analytically.  We integrate
the rest numerically. The absolute square of the scattering amplitude for
$\pi\rho\rightarrow e^+e^-$ (initial spin averaged
and final spin summed) is
\begin{eqnarray}
|\,\overline{\cal M}\,|^{2} &=& {4e^{4}\over 3} {|F_{\pi\rho}(M)|^{2}\over
M^{4}} \, \epsilon_{\mu\nu\alpha\beta}\, \epsilon_{\kappa\lambda
\sigma\tau}\, q^{\mu}q^{\kappa}p_{\rho}^{\alpha}p_{\rho}^{\sigma}
\left(-g^{\beta\tau}\right) \left[p_{+}^{\nu}p_{-}^{\lambda}
+ p_{-}^{\nu}p_{+}^{\lambda} -g^{\nu\lambda}(p_{+}\cdot p_{-})\right]
\label{eq:msquared1}
\end{eqnarray}
where where $q^{\mu}= p_{+}^{\mu}+p_{-}^{\mu}$ so that $q^{2}=M^{2}$
and finally,
where the form factor is given by Eq.~(\ref{eq:ffpirho}).

For later convenience it is useful to compute the cross section
arising from this squared amplitude.  We get
\begin{equation}
\sigma_{\pi\rho\to e^+e^-}(M) = {2\pi\alpha^{2}
\left|F_{\pi\rho}(M)\right|^{2}
\lambda^{1/2}(M^{2},m_{\pi}^{2},m_{\rho}^{2}) \over
9M^{2}}.
\end{equation}
Then the thermal rate can be written
\begin{equation}
{dR\over dM^{2}} = {{\cal N}\over 2(2\pi)^{4}} \int dE_{\pi}dE_{\rho}
f(E_{\pi})f(E_{\rho}) \sigma(s) \lambda^{1/2}(s,m_{\pi}^{2},m_{\rho}^{2})
\Theta(\chi)
\label{eq:newrate}
\end{equation}
where
\begin{equation}
\chi = 4\bbox{p}_{\pi}^{2}\bbox{p}_{\rho}^{2} -\left(m_{\rho}^{2}
+m_{\pi}^{2}-s+2E_{\rho}E_{\pi}\right)^{2}.
\end{equation}
Notice that $s$ and $M^{2}$ have
been used interchangeably.  The integration
required by Eq.~(\ref{eq:newrate}) can of course be done numerically for
arbitrary distributions.   However, assuming Boltzmann distributions
for the mesons it is a straigtforward exercise to perform the
integration and arrive at the much simpler expression for the rate
\begin{equation}
{dR\over dM^{2}} = {\cal N}\,{T\over 32\pi^{4}M}
\,K_{1}(M/T)\lambda(s,m_{\pi}^{2},m_{\rho}^{2})
\sigma(s)
\label{eq:rateclosed}
\end{equation}
where $K_{1}$ is the first modified Bessel function.
Making the replacements $m_{\rho}\rightarrow
m_{K^{*}}$, $m_{\pi}\rightarrow
m_{K}$ and using the appropriate degeneracy ${\cal N}$,
one arrives at the thermal rate for lepton pair emission through
$\bar{K}K^{*}+ {\rm c.c.} \rightarrow e^+e^-$.

Results at 150 MeV temperature presented in
Fig.~\ref{fig:pirhorate150}.  We have compared rates which were
integrated numerically starting from Bose-Einstein distributions
to those computed using classical distributions written finally as
Eq.~(\ref{eq:rateclosed}).  Since there is no final-state hadron
in these reactions and therefore no Bose-enhancement effects,
the classical distributions are more than adequate.
For comparison purposes we also include the results for $\pi^{+}\pi^{-}$
annihilation \cite{cgpl94} and for $\pi a_{1}$
reactions from Ref.~\cite{cscmkcg} which were concluded to
be dominant for masses above $m_{\phi}$.  The importance of the
excitations of $\omega(1390)$ and
$\phi(1680)$ are immediate: they are visible above the tail
of the $\pi^{+}\pi^{-}$ annihilation and even above the $\pi a_{1}$
results.  One last figure is shown (Fig.~\ref{fig:pirhorate200}) where
a temperature of 200 MeV is used.  Our conclusions are independent of
the temperature.

\section{Final Remarks}
\label{sec:conclusion}

Timelike photons are observed to have a strong affinity for neutral
hadronic fields.  This observation is captured in the current
field identity\cite{nk67} which is a consequence of the VMD hypothesis.
These photons, observed as lepton pairs, exhibit resonant structure
when the observed particles are projected onto a mass spectrum.  One
understands these peaks within vector-meson dominance to be
manifestations of hadronic form factors.  Our study suggests
that in order to improve knowledge of dielectron production rates
from hot hadronic matter, at least for dielectron masses ranging
from $m_{\phi}$ to $m_{J/\psi}$, form factors for hadrons
other than just the pion must be considered.  We show only two examples
of additional structure in the mass spectrum---there may well be more
surprises.

World data on electron-positron annihilation into
hadrons is quite extensive at higher energies but is unfortunately not
as complete for lower energies.  However, that which does exist
should be exploited to its fullest potential.  In our study we have
constructed effective form factors $|F_{\pi\rho}(M)|^{2}$ and
$|F_{KK^{*}}(M)|^{2}$ in the mass range 1--2.5 GeV.   After doing so,
we have compared each elementary process with known data before using
them in a thermal calculation on the extended system of interacting hadrons.
Higher excitations of the $\omega$ and $\phi$, namely $\omega(1390)$ and
$\phi(1680)$, were found to be visible in the mass distribution of the
dilepton production rate above the hadronic production mechanisms
previously considered.

\section*{Acknowledgments}

K.H. acknowledges support from the National
Science Foundation under grant number PHY-9403666 and wishes to
thank the Physics Department at McGill University for hospitality
during a visit in May 1994 when initial stages of this work were discussed.
C.G. acknowledges support from the Natural Sciences and Engineering
Research Council of Canada, the FCAR of the Qu\'ebec Government and a
NATO Collaborative Research Grant.

\begin{figure}
\caption{General annihilation diagram of two hadrons $a$ and $b$
to a lepton pair.  The process goes through a vector hadronic
resonance $V$ and by vector-meson dominance, goes directly
to a virtual photon.}
\label{fig:abtoe+e-}
\end{figure}
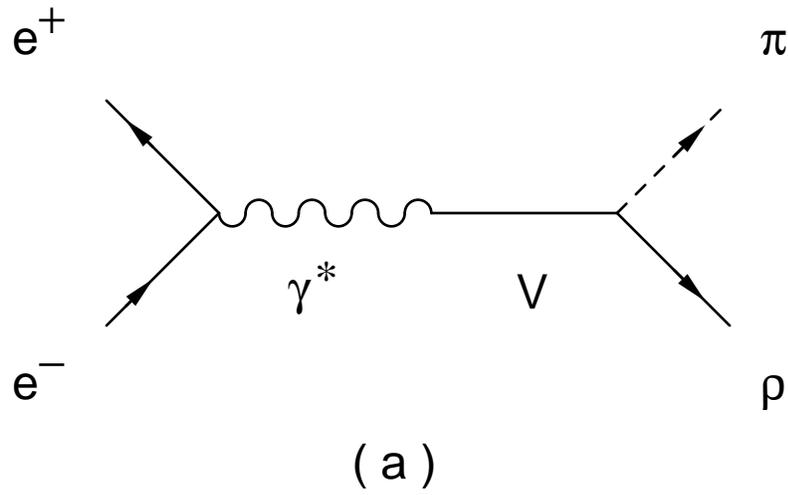
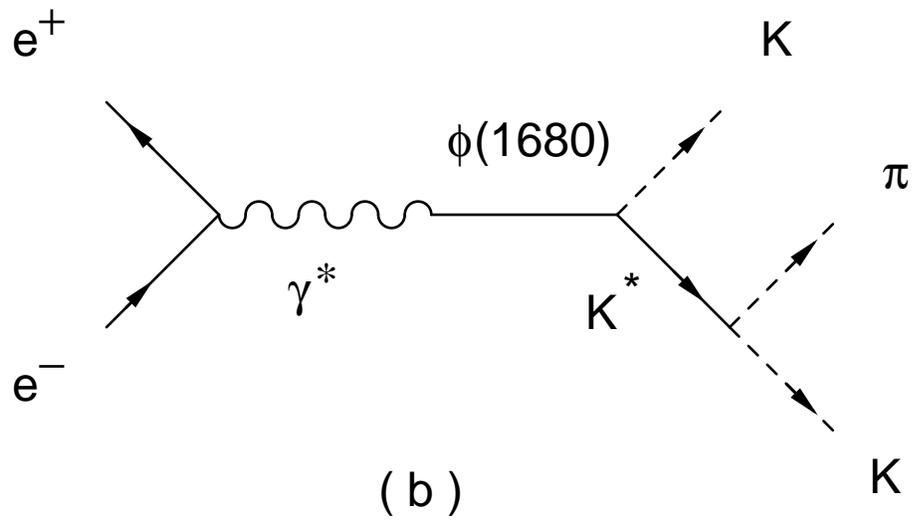
\begin{figure}
\caption{Annihilation of $e^+e^-$ into $\pi\rho$ through
any of three isoscalar vectors $\phi(1020)$, $\omega(1390)$ and
$\omega(1600)$ in (a) and in (b) $e^{+}e^{-} \to K_{S}^{0}K^{\pm}\pi^{\mp}$.}
\label{fig:timereversed}
\end{figure}
\begin{figure}
\caption{Cross section for $e^+e^-\to \pi\rho$
as computed within the model described in the text
as compared with experimental data from Ref.~\protect\cite{data1,data2}.}
\label{fig:e+e-topirho}
\end{figure}
\begin{figure}
\caption{Cross section for
$e^+e^-\to \bar{K}K^{*}$ described by the Breit-Wigner
form presented in the text (solid curve), results of a microscopic
calculation for the full process (dashed curve) and finally,
experimental data for the
reaction $e^+e^-\to K_{S}^{0}K^{\pm}\pi^{\mp}$
from Ref.~\protect\cite{data3}.}
\label{fig:e+e-tokkpi}
\end{figure}
\begin{figure}
\caption{Thermal production rate at $T=150$ MeV from $\pi\rho$ (solid
curve) processes, $\bar K K^{*} + $c.c. (dotted curve),
$\pi^{+}\pi^{-}$ annihilation from Ref.~\protect\cite{cgpl94} (dot-dash
curve) and finally, $\pi a_{1}$ process as calculated in
Ref.~\protect\cite{cscmkcg} (dashed curve).}
\label{fig:pirhorate150}
\end{figure}
\begin{figure}
\caption{Same as Fig.~\protect\ref{fig:pirhorate150} except using
$T=200$ MeV.}
\label{fig:pirhorate200}
\end{figure}


\begin{references}
%
\bibitem[*]{myemail} electronic address: haglin@theo03.nscl.msu.edu
\bibitem[\dag]{charlesemail} electronic address:
gale@hep.physics.mcgill.ca
%
\bibitem{cgpl94}C. Gale and P. Lichard, Phys. Rev. D {\bf 49}, 3338 (1994).
\bibitem{dm2coll}DM2 Collaboration, Phys. Lett. {\bf B 220}, 321 (1989).
\bibitem{cscmkcg}C. Song, C.M. Ko and C. Gale, Phys. Rev. D {\bf 50},
R1827 (1994).
\bibitem{khsp94}K. Haglin and S. Pratt, Phys. Lett. {\bf B 328}, 255 (1994).
\bibitem{pl94}P. Lichard, Phys. Rev. D {\bf 49}, 5812 (1994).
\bibitem{NA38}M. C. Abreu {\em et al.}, Nucl. Phys. {\bf A566},
77c (1994); and references therein.
\bibitem{um88}U.-G. Meissner, Phys. Rep. {\bf 161}, 213 (1988).
\bibitem{js60}J.J. Sakurai, Ann. Phys. (N.Y.) {\bf 11}, 1 (1960).
\bibitem{khcg94}K.L. Haglin and C. Gale, Nucl. Phys. {\bf B421}, 613 (1994).
\bibitem{data1}V.M. Aulchenko {\em et al.}, Novosibirsk Preprint 86--106
(1986); A. Donnachie and A.B. Clegg, Z. Phys. C {\bf 42}, 663 (1989).
\bibitem{data2}R. Baldini--Ferroli in Proc. Had. Phys. at Intermediate
Energy, T. Bressani, B. Menetti, and G. Pauli (eds.) Amsterdam, New York,
Elsevier (1987); A. Donnachie and A.B. Clegg, Z. Phys. C {\bf 42}, 663 (1989).
\bibitem{data3}F. Mane, {\em et al.}, Phys. Lett. {\bf B 112}, 179 (1982).
\bibitem{nk67}N. Kroll, T.D. Lee and B. Zumino, Phys. Rev. {\bf 157}, 1376
(1967).
%
\end{references}
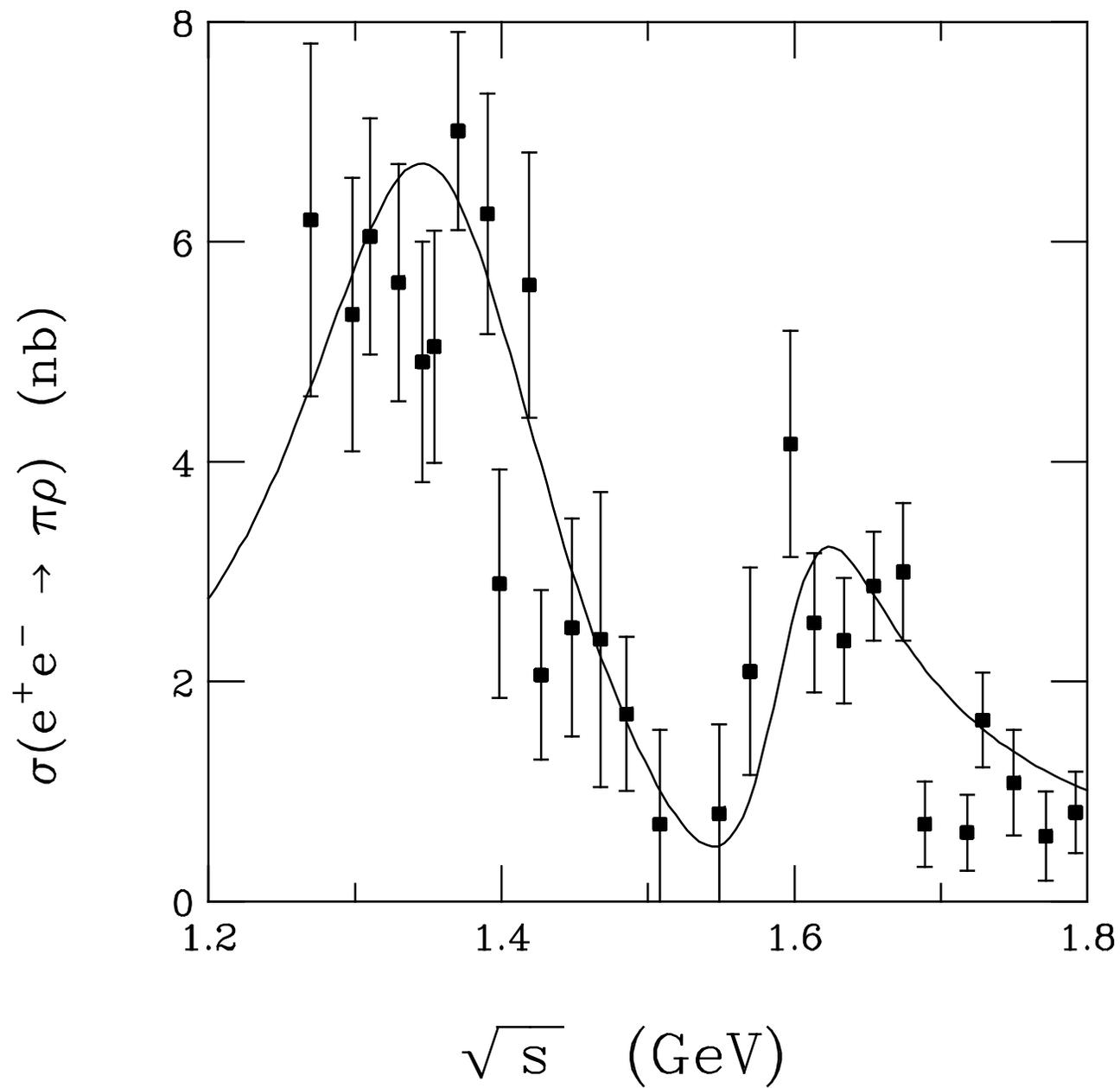
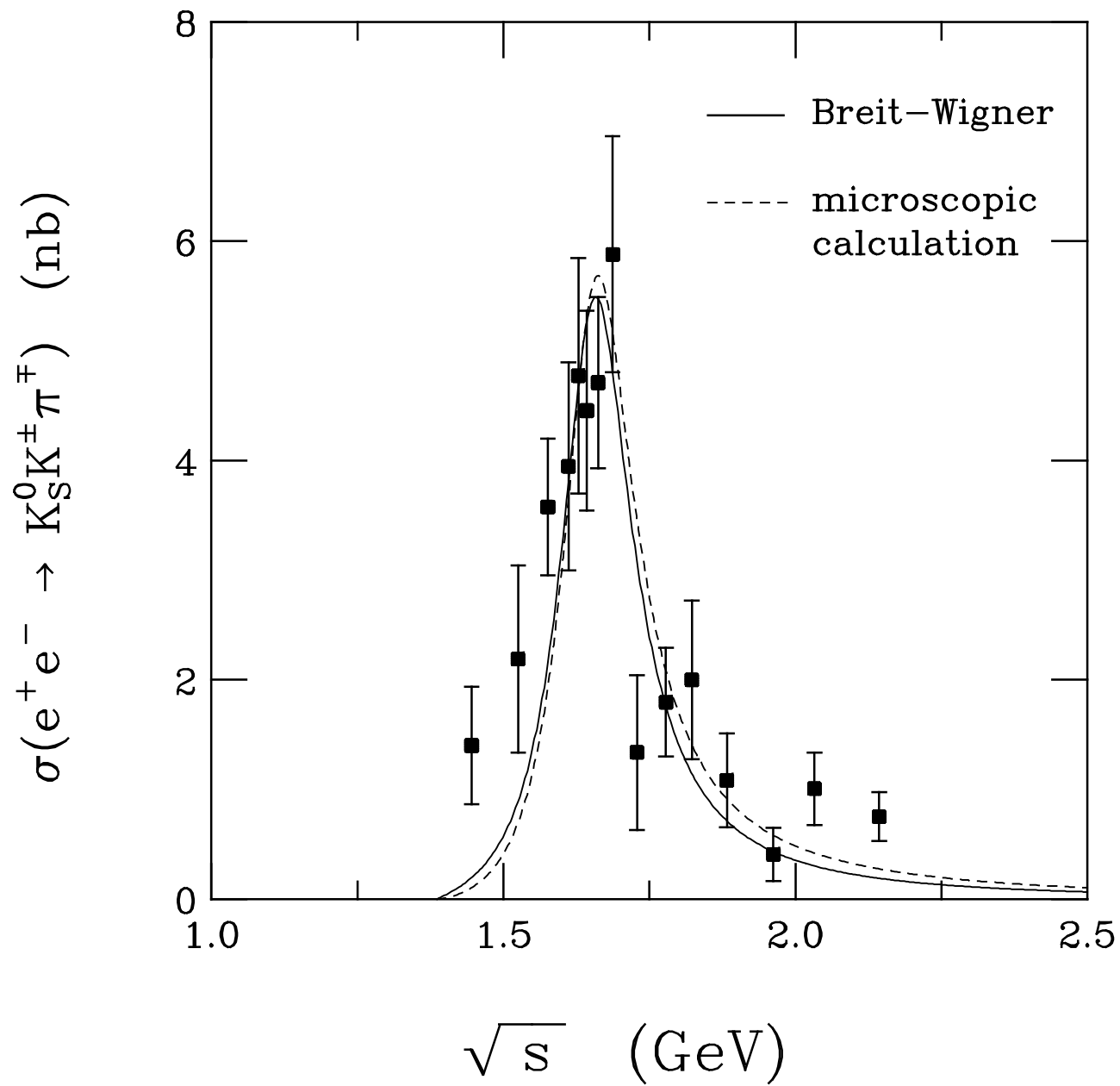
\end{document}